\documentclass[12pt,twoside,a4paper,fleqn]{article}
\usepackage[left=25mm,top=20mm,right=14mm,bottom=25mm]{geometry}
\usepackage{epsfig}
\usepackage{graphicx}
\usepackage{amssymb}
\usepackage{mathrsfs}
\usepackage{dcolumn}
\usepackage{amsmath}
\usepackage[utf8]{inputenc}
\usepackage[T1]{fontenc}
\usepackage{float}
\usepackage{multirow}
\usepackage{hyperref}
\usepackage{appendix}
\newcommand{\pr}{\partial}
\newcommand{\rta}{\rightarrow}

\newcommand{\upa}{\uparrow}
\newcommand{\dwa}{\downarrow}
\newcommand{\ep}{\epsilon}

\newcommand{\p}{\prime}
\newcommand{\om}{\omega}
\newcommand{\ra}{\rangle}
\newcommand{\la}{\langle}

\newcommand{\beq}{\begin{equation}}
	\newcommand{\eeq}{\end{equation}}

\newcommand{\ball}{\begin{align}}
	\newcommand{\eall}{\end{align}}

\newcommand{\beqar}{\begin{eqnarray}}
	\newcommand{\eeqar}{\end{eqnarray}}

\newcommand{\dg}{\dagger}
\newcommand{\mud}{\mu_{d}}
\newcommand{\mus}{\mu_{s}}
\newcommand{\lam}{\lambda}
\newcommand{\omp}{\om^\p}

\makeatletter
\newcommand*{\rom}[1]{\expandafter\@slowromancap\romannumeral #1@}
\makeatother

\begin{document}
	\date{}
	\title{Non-Drude behaviour of optical conductivity in Kondo-lattice systems}
	\author{Komal Kumari\footnote{Visiting Physical Research Laboratory, Ahmedabad}\\ 
		Department~of~Physics,~Himachal~Pradesh~University,\\~Shimla,~India, Pin:171005.}
	\maketitle
	\begin{abstract}
		The  optical conductivity in a Kondo lattice system is presented in terms of the  memory function formalism. I use Kondo-lattice Hamiltonian for explicit calculations. I compute the frequency dependent  imaginary part of the memory function ($M^{\p\p}(\om)$), and the  real part of the memory function $M^{\p}(\om)$ by using the Kramers-Kronig transformation. Optical conductivity is computed using the generalized Drude formula. I find that high frequency tail of the optical conductivity scales as $\sigma(\om) \sim \frac{1}{\om}$ instead of the Drude $\frac{1}{\om^2}$ law. Such a  behaviour is seen in strange metals. My work points out that it may be the magnetic scattering mechanisms that are important for the anomalous behaviour of strange metals.
	\end{abstract}
\section{Introduction}
In the Drude model, optical conductivity is given by \cite{ashcroft1976solid}
\beqar
\sigma(\om)= \frac{n e^2}{m} \frac{1}{i\om+\frac{1}{\tau_{D}}}. \label{d1}
\eeqar
here $\tau_{D}$ is the Drude scattering rate. The real and imaginary parts of the conductivity are given by
\beq
\sigma^{\p}(\om)=\frac{n e^2}{m}\frac{\frac{1}{\tau_{D}}}{\om^2+(\frac{1}{\tau_{D}})^2}. \label{dr2}
\eeq
\beq
\sigma^{\p\p}(\om)=\frac{n e^2}{m}\frac{\om}{\om^2+(\frac{1}{\tau_{D}})^2}. \label{dr3} 
\eeq
In the high frequency limit $\om\gg \frac{1}{\tau_{D}}$, the real part of conductivity scales $\sigma^{\p}(\om)\sim\frac{1}{\om^2}$. This is a typical signature of "good" metals \cite{ashcroft1976solid,PhysRevB.6.1226}. But it has been experimentally observed that $\sigma^{\p}(\om)$ scales as $\frac{1}{\om}$ or some fractional power of $\frac{1}{\om}$ in many situations (such as in the "normal" state of high temperature superconductors).The metallic states which show such a behaviour are called strange metals \cite{Singh2017}\\

The aim of the present work is to show that the behaviour  $\sigma^{\p}(\om)\sim \frac{1}{\om}$ can arise in a situation where electron scattering happens via magnetic spin fluctuations, whereas the standard electron-impurity and electron-phonon scaterring lead to the Drude behaviour ($\sigma^{\p}(\om)\sim \frac{1}{\om^2}$).

\section{Drude Theory and its Generalizations}
 In simple Drude model a frequency independent time $\tau_{D}$ governs the relaxation of current and $\frac{1}{\tau_{D}}$ is identified as the Drude scattering rate. The simple Langevin equation leads to \cite{PhysRevB.6.1226,Singh2017,komal_nav}
\beq
\sigma(\om) = \frac{ne^2}{m} \frac{1}{i\om +\frac{1}{\tau_{D}}}  \label{ac1}
\eeq
Here $n$ and $m$ are the number density and mass of the free electrons. The standard Drude formula has many limitations \cite{komal_nav}. The  Generalized Drude Formula (GDF) writes the dynamical conductivity  in terms of the memory function\cite{PhysRevB.6.1226,Singh2017,komal_nav}
\beq
\sigma(\om) = \frac{n e^2}{m} \frac{1}{i\om +M(\om)}.
\label{gdf1}
\eeq
where $M(\om)$ is the complex frequency dependent memory function. In some situations when $M(\om)$ becomes the frequency independent, the generalized Drude formula (\ref{gdf1}) reduces to simple Drude formula(\ref{ac1})
\cite{PhysRevB.6.1226,Singh2017,komal_nav}.
\subsection{\Large $A.C.$ Conductivity}
In this section I write the dynamical conductivity by introducing $\sigma(z)$ in terms of memory function $M(z)$ by introducing complex frequency $z$ \cite{Singh2017}.
\beqar
\sigma(z)=i\frac{ne^2}{m}\frac{1}{z+M(z)}, \label{sig1}
\eeqar
here complex function $M(z=\om\pm i0)$ can be written as $M^{\p}(\om)\pm  i M^{\p\p}(\om)$. Therefore $\sigma(z)$ can be separated into real $\sigma(\om \pm i0)$ and imaginary $\sigma^{\p\p}(\om)$ parts: 

\beqar
\sigma^{\p}(\om)= \frac{ne^2}{m} \frac{ M^{\p\p}(\om)}{(\om+M^{\p}(\om))^2+(M^{\p\p}(\om))^2}, \label{realcond1}
\eeqar 
\beqar\sigma^{\p\p}(\om)= \frac{ne^2}{m} \frac{( \om+M^{\p}(\om))}{(\om+M^{\p}(\om))^2+(M^{\p\p}(\om))^2}.  \label{imgcond1}
\eeqar 
My next task is to compute the frequency dependent real and imaginary parts of the memory function and then dynamical conductivity can be computed.
\section{The memory function for $s$ -$d$ Hamiltonian}
I take the  $s$-$d$ Hamiltonian (known as Kondo-lattice Hamiltonian) for explicit calculations, and compute the imaginary part of the memory function. The Kondo-lattice Hamiltonian is given by

\beqar
H_{sd}&=&\frac{J}{N}\sum_{k^{\p}k} \bigg\{ c^{\dagger}_{k^{\p}\uparrow} c_{k\downarrow}S^{-}(k^{\p}-k)+ c^{\dagger}_{k^{\p}\downarrow} c_{k\uparrow}S^{+}(k^{\p}-k)+(c^{\dagger}_{k^{\p}\uparrow}
c_{k\upa}-c^{\dagger}_{k^{\p}\downarrow} c_{k\dwa})S^{z}(k^{\p}-k)\bigg\}  \label{sd1}
\eeqar
here $J$ is  coupling constant between $s$-electron and $d$ electrons \cite{komal_2020}.   $c^\dg$ and $c$ are the creation and annihilation operators for $s$-electrons.   $S^{-}(k^{\p}-k)$ and $S^{+}(k^{\p}-k)$ are the spin lowering and spin raising  operator of $d$ or $f$ electrons and these are given as
\beqar 
S^{-}(q)=\sum_{k}a^*_{k+q\dwa}a_{k\upa},~~~~~~~~~S^{+}(q)=\sum_{k}a^*_{k+q\upa}a_{k\dwa}. \label{spind1}
\eeqar 
	Under the assumption of weak coupling (electron K.E.$\gg J$) between the $s$ and $d$ electrons the W\"{o}lfle-G\"{o}tze equation of motion method defines the memeory function as \cite{komal_2020,rani2017,bhalla2016,PhysRevbhalla}
\beqar
M(z)\simeq \frac{1}{z}(\frac{m}{ne^2})[\la\la \dot{J_{1}};\dot{J_{1}}\ra\ra_{z}-\la\la \dot{J_{1}};\dot{J_{1}}\ra\ra_{0}] \label{eq3}.
\eeqar
where $ \dot{J_{1}}=-\frac{i}{\hbar}[J_{1},H_{sd}]$.
 $J_{1}=\frac{1}{V}\sum_{k\sigma}ev_{k}c^{\dagger}_{k\sigma}c_{k\sigma}$ is the current density operator and $V$ is volume of the sample. I define the current-current correlator as $\phi(z)=\la\la\dot{J_{1}};\dot{J_{1}}\ra\ra$ and compute the correlator in terms of fermi functions of electrons(for more details refer to \cite{komal_2020}), we get
  \beqar
 \phi(z)&=&-\frac{e^2J^2}{N^2\hbar^3 V}\sum_{k^{\p}k}(v_{1}(k^\p)-v_{1}(k))^2\bigg\{ f^s_{k^\p}(1-f^s_{k}) \sum_{k_{d},k^{\p}_{d}} (f^{d}_{k_{d}}-f^{d}_{k^{\p}_{d}})-(f^s_{k}-f^s_{k^\p}) \nonumber\\&&\sum_{k_{d},k^{\p}_{d}} f^{d}_{k_{d}}(1-f^d_{k^{\p}_{d}})\bigg\}  
 ~\bigg[ \frac{1}{\frac{\ep_{k^\p}}{\hbar}-\frac{\ep_{k}}{\hbar}-\om_{k^\p-k}+z} +\frac{1}{\frac{\ep_{k^\p}}{\hbar}-\frac{\ep_{k}}{\hbar}-\om_{k^\p-k}-z}  \bigg]. \label{c9}
 \eeqar
On substituting the expression (\ref{c9}) in (\ref{eq3}) the frequency depedent memory function can be expressed as 
\beqar
M(z)&=&-\frac{J^2 m}{N^2\hbar^3 n  V \om}\sum_{k^{\p}k}(v_{1}(k^\p)-v_{1}(k))^2\bigg\{ f^s_{k^\p}(1-f^s_{k})\sum_{k_{d},k^{\p}_{d}} (f^{d}_{k_{d}}-f^{d}_{k^{\p}_{d}})- (f^s_{k}-f^s_{k^\p})\nonumber\\&& \sum_{k_{d},k^{\p}_{d}}f^{d}_{k_{d}}(1-f^d_{k^{\p}_{d}})\bigg\}   
\bigg[ \frac{1}{\frac{\ep_{k^\p}}{\hbar}-\frac{\ep_{k}}{\hbar}-\om_{k^\p-k}+z} +\frac{1}{\frac{\ep_{k^\p}}{\hbar}-\frac{\ep_{k}}{\hbar}-\om_{k^\p-k}-z}\nonumber\\&&~~~~~~~~~~~~~-\frac{1}{\frac{\ep_{k^\p}}{\hbar}-\frac{\ep_{k}}{\hbar}-\om_{k^\p-k}}-\frac{1}{\frac{\ep_{k^\p}}{\hbar}-\frac{\ep_{k}}{\hbar}-\om_{k^\p-k}} \bigg], \label{me1}
\eeqar 
here we write the short notation for $f^1_{d}(q)=\sum_{k_{d},k^{\p}_{d}} (f^{d}_{k_{d}}-f^{d}_{k^{\p}_{d}})$ and $f^1_{d}(q)=\sum_{k_{d},k^{\p}_{d}}f^{d}_{k_{d}}(1-f^d_{k^{\p}_{d}})$.
From the above expression the imginary part of the memory function can be written as (for more details refer to ref. \cite{komal_2020})
\begin{multline}
	M^{\p\p}(\om)=\frac{1}{12\pi^3}\frac{J^2 V m^2 }{N^2\hbar^6  n q^2_{s} }\int_{0}^{q_{D}}dq q^3 \bigg[ \int_{0}^{\infty}d\ep\sqrt{\ep}\times\\ \bigg(\underbrace{\frac{\sqrt{\ep+\hbar\om_{q}-\hbar\om}f^s(\ep+\hbar\om_{q}-\hbar\om)-\sqrt{\ep+\hbar\om_{q}+\hbar\om}f^s(\ep+\hbar\om_{q}+\hbar\om)}{\om}}_{\mathfrak{T}_{1}}\bigg)\\~\times(1-f^s(\ep))f^1_{d}(q)+\int_{0}^{\infty}d\ep\sqrt{\ep} \bigg(\underbrace{\frac{\sqrt{\ep+\hbar\om_{q}+\hbar\om}-\sqrt{\ep+\hbar\om_{q}-\hbar\om}}{\om}}_{\mathfrak{T}_{2}}\bigg)\times\\f^s(\ep)f^2_{d}(q)+ \int_{0}^{\infty}\frac{d\ep\sqrt{\ep}}{\om} \bigg(\sqrt{\ep+\hbar\om_{q}-\hbar\om}f^s(\ep+\hbar\om_{q}-\hbar\om)-~~~~~~\\~\sqrt{\ep+\hbar\om_{q}+\hbar\om}f^s(\ep+\hbar\om_{q}+\hbar\om)\bigg)f^2_{d}(q)\bigg].~~~~~  \label{m11}
\end{multline}
In the next section I compute the frequency dependent part of the memory function under the assumption of the long wavelength limit.

\section{\Large Computation of frequency dependent Memory Function}
\begin{figure}
	\begin{tabular}{ccc}
		\includegraphics[height = 4.5cm, width =6.5cm]{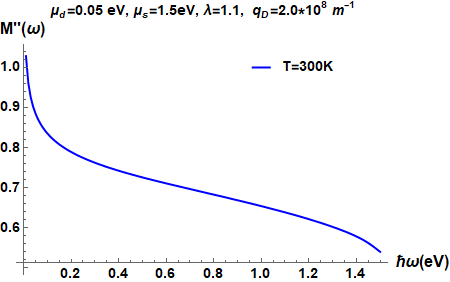}&~~~~
		\includegraphics[height = 4.5cm, width =6.5cm]{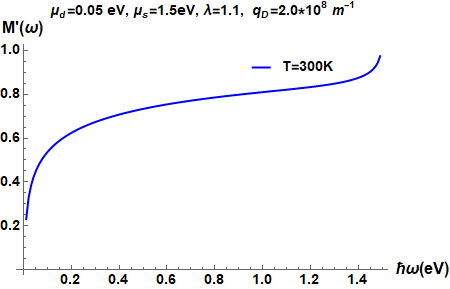}\\
		(a)&
		(b)\\
	\end{tabular}
	\caption{(a,b) Present the real and imaginary parts of the memory function.}
	\label{figchap4:aa}
\end{figure}

\begin{figure}{}
	\centering
	\includegraphics[height = 6.5cm, width =10cm]{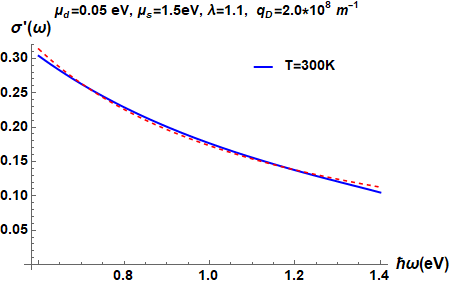}
	\caption{Represents the least square fit for tail part of the real part of conductivity. Red dashed curve corresponds to the least square fit: $\sigma=-0.03795 + \frac{0.2111}{\om}$. Thus we conclude that at higher frequencies $\sigma(\om) \sim \frac{1}{\om}$.}
	\label{chapfig4:b}
\end{figure}

I assume that momentum randomization of $s$-electrons happens via the creation of magnetic spin waves in the sub-system of $d$-electrons. This is the mechanism  of resistivity in the considered setting. In equation (\ref{m11}) $\hbar\om_{q}$ is the energy of the magnetic spin waves. One  takes $\hbar\om_{q}=c_{m}q^2$, where $c_{m}$ is a constant. I further assumes that ($\hbar\om_{q_{D}}\ll \mus$). That is the maximum energy of magnetic spin waves is much less than the chemical potential of $s$-electrons. Under these assumptions terms $\mathfrak{T}_{1}$ and $\mathfrak{T}_{2}$ in equation (\ref{m11}) can be approximated as :
\begin{multline}
	\mathfrak{T}_{1}\simeq\frac{\sqrt{\ep-\hbar\om}}{e^{\beta(\ep-\hbar\om-\mus)}+1}-\frac{\sqrt{\ep+\hbar\om}}{e^{\beta(\ep+\hbar\om-\mus)}+1}+\frac{q^2c_{m}}{2}\bigg( \frac{1}{\sqrt{\ep-\hbar\om}(1+e^{\beta(\ep-\hbar\om-\mus)})}\\~~~~-\frac{1}{\sqrt{\ep+\hbar\om}(1+e^{\beta(\ep+\hbar\om-\mus)})}-2\beta\frac{\sqrt{\ep-\hbar\om}e^{\beta(\ep-\hbar\om-\mus)}}{(1+e^{\beta(\ep-\hbar\om-\mus)})^2}+2\beta\frac{\sqrt{\ep+\hbar\om}e^{\beta(\ep+\hbar\om-\mus)}}{(1+e^{\beta(\ep+\hbar\om-\mus)})^2} \bigg)+O(q^4).~~~~~~ \label{m2}
\end{multline}
\beqar
\mathfrak{T}_{2}&\simeq&\sqrt{\ep+\hbar\om}-\sqrt{\ep-\hbar\om}+\frac{q^2c_{m}}{2} \bigg( \frac{1}{\sqrt{\ep+\hbar\om}}-\frac{1}{\sqrt{\ep-\hbar\om}}\bigg)+O(q^4).\nonumber\\&& \label{m3}
\eeqar
On Substituting expanded form of the terms  $\mathfrak{T}_{1}$ and $\mathfrak{T}_{2}$ in expression (\ref{m11}), we obtain
\begin{multline}
	M^{\p\p}(\om)=\frac{J^2Vm^2}{12\pi^3N^2\hbar^6n q^2_{s}}\int_{0}^{q_{D}}dq q^3 \bigg[ \int_{0}^{\infty}\frac{d\ep\sqrt{\ep}}{\om}\bigg\{\frac{\sqrt{\ep-\hbar\om}}{(1+e^{\beta(\ep-\hbar\om-\mus)})}-\frac{\sqrt{\ep+\hbar\om}}{(1+e^{\beta(\ep+\hbar\om-\mus)})}+\\~\frac{q^2c_{m}}{2}\bigg(\frac{(\sqrt{\ep-\hbar\om})^{-1}}{(1+e^{\beta(\ep-\hbar\om-\mus)})}-\frac{(\sqrt{\ep+\hbar\om})^{-1}}{(1+e^{\beta(\ep+\hbar\om-\mus)})}-2\beta\frac{\sqrt{\ep-\hbar\om}e^{\beta(\ep-\hbar\om-\mus)}}{(1+e^{\beta(\ep-\hbar\om-\mus)})^2}+2\beta\frac{\sqrt{\ep+\hbar\om}e^{\beta(\ep+\hbar\om-\mus)}}{(1+e^{\beta(\ep+\hbar\om-\mus)})^2}\bigg)+~\\~~O(q^4)
	\bigg\} (1-f^{s}(\ep))f^{1}_{d}(q)+\int_{0}^{\infty}\frac{d\ep\sqrt{\ep}}{\om}\bigg\{\sqrt{\ep+\hbar\om}-\sqrt{\ep-\hbar\om}+\frac{q^2c_{m}}{2}\bigg(\frac{1}{\sqrt{\ep+\hbar\om}}- \frac{1}{\sqrt{\ep-\hbar\om}}\bigg)\bigg\}~\\~f^{s}(\ep)f^{2}_{d}(q)+\int_{0}^{\infty}\frac{d\ep\sqrt{\ep}}{\om}\bigg\{\frac{\sqrt{\ep-\hbar\om}}{(1+e^{\beta(\ep-\hbar\om-\mus)})}-\frac{\sqrt{\ep+\hbar\om}}{(1+e^{\beta(\ep+\hbar\om-\mus)})}+\frac{q^2c_{m}}{2}\bigg(\frac{(\sqrt{\ep-\hbar\om})^{-1}}{(1+e^{\beta(\ep-\hbar\om-\mus)})}-~\\~\frac{(\sqrt{\ep+\hbar\om})^{-1}}{(1+e^{\beta(\ep+\hbar\om-\mus)})}-2\beta\frac{\sqrt{\ep-\hbar\om}e^{\beta(\ep-\hbar\om-\mus)}}{(1+e^{\beta(\ep-\hbar\om-\mus)})^2}+2\beta\frac{\sqrt{\ep+\hbar\om}e^{\beta(\ep+\hbar\om-\mus)}}{(1+e^{\beta(\ep+\hbar\om-\mus)})^2}\bigg)+O(q^4)\bigg\}f^{2}_{d}(q,\ep_{d})\bigg],~~~~~~~~~~~~\label{m4}
\end{multline}
my next task is to perform $q$ integration in the expression (\ref{m4}). The terms $f^{1}_{d}(q)$  and $f^{2}_{d}(q)$ also contribute $q$ terms in the integrals. The expansion of  $f^{1}_{d}(q)=\sum_{k_{d}}[ f^{d}(\ep_{k_{d}})-f^d(\ep_{k^{\p}_{d}})]$ in long wavelength limit ($q\rta0$) gives
\beqar
f^1_{d}(q)&=& \sum_{k_{d}}[ f^{d}(\ep_{k_{d}})- f^{d}(\ep_{k_{d}})-q\frac{\pr f^{d}(\ep_{k^{\p}_{d}})}{\pr q}|_{q=0}-\frac{q^2}{2!}\frac{\pr^2 f^{d}(\ep_{k^{\p}_{d}})}{\pr q^2}|_{q=0}-\frac{q^3}{3!}\frac{\pr^3 f^{d}(\ep_{k^{\p}_{d}})}{\pr q^3}|_{q=0}..]. \nonumber\\ \label{fd1}
\eeqar
On converting summation into integrals ($\sum_{k_{d}}=\frac{V}{(2\pi)^3}\int d^3k_{d}$), we get
\beqar
f^1_{d}(q)&=&-\frac{V}{(2\pi)^2}\int_{0}^{\infty}k^2_{d} dk_{d}\int_{0}^{\pi}\sin\theta d\theta\bigg[q\frac{\pr f^{d}(\ep_{k^{\p}_{d}})}{\pr q}|_{q=0}+\frac{q^2}{2!}\frac{\pr^2 f^{d}(\ep_{k^{\p}_{d}})}{\pr q^2}|_{q=0}\nonumber\\&&~~~~~~~~~~~~~~~~~+\frac{q^3}{3!}\frac{\pr^3 f^{d}(\ep_{k^{\p}_{d}})}{\pr q^3}|_{q=0}.......\bigg].\label{fd2}
\eeqar
Further simplification of $f^1_{d}(q)$,  we obtain (for more details refer to 
\cite{komal_2020})
\begin{multline}
	f^1_{d}(\ep_{d},q)=V\frac{q^2}{4\pi^2}\frac{\sqrt{2m_{d}}}{\hbar}\bigg[\beta \int_{0}^{\infty}\frac{d\ep_{d} \sqrt{\ep_{d}}e^{\beta(\ep_{d}-\mu_{d})}}{(e^{\beta(\ep_{d}-\mu_{d})}+1)^2}+\frac{2}{3}\beta^2 \int_{0}^{\infty} \frac{d\ep_{d} \ep_{d}^{\frac{3}{2}}e^{\beta(\ep_{d}-\mu_{d})}}{(e^{\beta(\ep_{d}-\mu_{d})}+1)^2}-\\~~~~~~~~~~~~~~\frac{4}{3}\beta^2\int_{0}^{\infty} \frac{d\ep \ep^{\frac{3}{2}}e^{2\beta(\ep_{d}-\mu_{d})}}{(e^{\beta(\ep_{d}-\mu_{d})}+1)^3}\bigg]. \label{fd7}
\end{multline}
Let $m_{d}$ be the mass of the $d$-electrons and m is the mass of $s$-electrons, introduces $m_{d}=m\lambda$ and set $\sqrt{2m\mus}=\hbar q_{s}$
The expression $f^1_{d}(\ep_{d})$ can be written as 
\begin{multline}
	f^1_{d}(\ep_{d},q)= V q_{s} \frac{q^2}{4 \pi^2}\sqrt{ \frac{\lam}{\beta\mus}}\bigg(\int_{-\beta\mud}^{\infty}\frac{dx \sqrt{x+\beta\mud} e^{x}}{(e^{x}+1)^2}+\frac{2}{3}\int_{-\beta\mud}^{\infty}\frac{dx (x+\beta\mud)^{\frac{3}{2}} e^{x}}{(e^{x}+1)^2}-\\~~\frac{4}{3}\int_{-\beta\mud}^{\infty}\frac{dx (x+\beta\mud)^{\frac{3}{2}} e^{2x}}{(e^{x}+1)^3}\bigg),~~~~~~~ \label{fd8}
\end{multline}
and on the similiar lines the $f^2_{d}(\ep_{d},q)$ expression gives
\begin{multline}
	f^2_{d}(\ep_{d},q)=  \frac{V q^3_{s}~}{4 \pi^2}(\frac{\lam}{\beta\mus})^{\frac{3}{2}} \int_{-\beta\mud}^{\infty}\frac{dx \sqrt{x+\beta\mud} e^{x}}{(e^{x}+1)^2}+ \frac{V q_{s} q^2}{4 \pi^2}\sqrt{ \frac{\lam}{\beta\mus}}\\~\bigg(\int_{-\beta\mud}^{\infty}\frac{dx \sqrt{x+\beta\mud} e^{x}}{(e^{x}+1)^3}+\frac{2}{3}\int_{-\beta\mud}^{\infty}\frac{dx (x+\beta\mud)^{\frac{3}{2}} e^{x}}{(e^{x}+1)^3}-\frac{4}{3}\int_{-\beta\mud}^{\infty}\frac{dx (x+\beta\mud)^{\frac{3}{2}} e^{2x}}{(e^{x}+1)^4}\bigg).~~~~~ \label{fd9}
\end{multline}
On substitution the simplified form of the expressions $f^1_{d}(\ep_{d},q)$ and $f^2_{d}(\ep_{d},q)$  from equation (\ref{fd8}) and (\ref{fd9}) in the expression (\ref{m4}) and computing $q$ integral, we get
\begin{multline}
	M^{\p\p}(\om)=\frac{J^2Vm^2}{12\pi^3N^2\hbar^8n} \bigg[ \sqrt{\frac{\lam}{\beta\mus}} \int_{0}^{\infty}\frac{d\ep\sqrt{\ep}}{\om}\bigg((\frac{q_{D}}{q_{s}})^6\frac{q^5_{s}}{6} h_{1}(\ep,\hbar\om)+(\frac{q_{D}}{q_{s}})^{8} \frac{c_{m}q^7_{s}}{16}h_{2}(\ep,\hbar\om)\bigg)
	\\	\times (1-f^{s}(\ep)) \Xi (\beta\mud)+(\frac{\lam}{\beta\mus})^{\frac{3}{2}}  \int_{0}^{\infty}\frac{d\ep\sqrt{\ep}}{\om}\bigg( (\frac{q_{D}}{q_{s}})^4\frac{q^5_{s}}{4}h_{3}(\ep,\hbar\om)+ (\frac{q_{D}}{q_{s}})^6\frac{c_{m} q^7_{s}}{12}h_{4}(\ep,\hbar\om)\bigg)f^{s}(\ep)\\~\Xi_{1} (\beta\mud)+~\sqrt{\frac{\lam}{\beta\mus}}\int_{0}^{\infty}\frac{d\ep\sqrt{\ep}}{\om}\bigg((\frac{q_{D}}{q_{s}})^6\frac{q^5_{s}}{6}h_{3}(\ep,\hbar\om)+ (\frac{q_{D}}{q_{s}})^8\frac{c_{m} q^7_{s}}{16}h_{4}(\ep,\hbar\om)\bigg)f^{s}(\ep)\Xi_{2}(\beta\mud)+\\~(\frac{\lam}{\beta\mus})^{\frac{3}{2}}\times \int_{0}^{\infty}\frac{d\ep\sqrt{\ep}}{\om}\bigg((\frac{q_{D}}{q_{s}})^4\frac{q^5_{s}}{4}h_{1}(\ep,\hbar\om)+ (\frac{q_{D}}{q_{s}})^6\frac{c_{m} q^5_{s}}{12}h_{2}(\ep,\hbar\om)\bigg)\Xi_{1}(\beta\mud)+~~\\~ \sqrt{\frac{\lam}{\beta\mus}}\times\int_{0}^{\infty}\frac{d\ep\sqrt{\ep}}{\om}\bigg( (\frac{q_{D}}{q_{s}})^6\frac{q^5_{s}}{6} h_{1}(\ep,\hbar\om)+(\frac{q_{D}}{q_{s}})^8\frac{c_{m}q^7_{s}}{16}h_{2}(\ep,\hbar\om)\bigg)\Xi_{2}(\beta\mud)\bigg]. ~~~~~~\label{fd12}
\end{multline}
Here the terms $h_{1}(\ep,\hbar\om)$, $h_{2}(\ep,\hbar\om)$, $h_{3}(\ep,\hbar\om)$,	$h_{4}(\ep,\hbar\om)$, $\Xi $, $\Xi_{1}$, and  $\Xi_{2}$ are defined in Appendix. The above obtained result  is the final expression of the frequency dependent imaginary part of the memory function. I will numerically compute $M^{\p\p}(\om)$ for certain values of  parameters $\mud$, $\mus$,(chemical potential of $d$ and $s$ electrons) $q_{D}$ , $q_{s}$ and $\lam$. To find $A.C.$  conductivity one also needs the real part of memory function, which is computed by employing the Kramers–Kronig relation \cite{PhysRevB.84.045429,Dynamical1990,PhysRevB.69.214514,lucarini2005kramers,PhysRevB.87.115109}:
\beqar
M^{\p}(\om)=-\frac{2}{\pi}\int d\omp \frac{ \omp M^{\p\p}(\omp) } {\om^{\p 2}-\om^2}.
\eeqar
In figure \ref{figchap4:aa} I plot the $M^{\p\p}(\om)$ and $M^{\p}(\om)$ for the values of the fitting parameters which are writtem in the head caption of the figure.
\floatplacement{figure}{H} 
Next, I numerically calculate the $A.C.$ conductivity using equations (\ref{realcond1} and \ref{imgcond1}).
\floatplacement{figure}{H}
\begin{figure}
	\begin{tabular}{ccc}
		\includegraphics[height = 4.5cm, width =6.5cm]{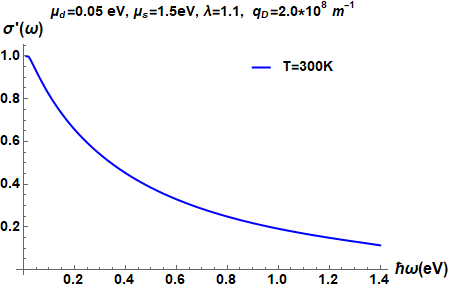}&~~~~
		\includegraphics[height = 4.5cm, width =6.5cm]{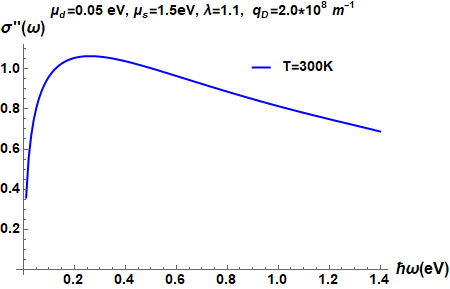}\\
		(a)&
		(b)\\
	\end{tabular}
	\caption{(a,b) Present the real and imaginary parts of $A.C.$ conductivity. }
	\label{figchap4:a}
\end{figure}
Figure \ref{figchap4:a} presents the $A.C.$ conductivity behaviour of the Kondo-lattice system. Figure \ref{figchap4:a} (a) presents the real part of the $A.C.$ conductivity for the values of the fitting parameters $\mud=0.05eV, \mus=1.5eV, \lam=1.1$ and $q_{D}=2.0\times  10^8 m^{-1}$. In high frequency regime $\frac{1}{\om}$ fitting for the real part of conductivity is performed using the the least square fitting method.  Figure \ref{chapfig4:b} shows least square fitting in high frequency regime (tail part of figure \ref{chapfig4:b}). Red dashed line shows $\frac{1}{\om}$ comparison. Thus, it is found that  the tail part of the real conductivity scales to $\frac{1}{\om}$. In real $\sigma(\om)$ at lower frequencies we observe the Drude peak and at higher $\om$ we observe that $\sigma(\om) \sim \frac{1}{\om}$ (instead of $\frac{1}{\om^2}$ Drude law) \cite{Singh2017,PhysRevB.84.045429}. Figure \ref{chapfig4:b} shows least square fitting in high frequency regime (tail part of figure \ref{chapfig4:b}). Red dashed line shows $\frac{1}{\om}$ comparison.
\section{Conclusion}
The calculations of $A.C.$ conductivity using the memory function formalism for the Kondo-lattice Hamiltonian is presented. Using the W\"{o}lfle-G\"{o}tze  equation of motion method  the frequency dependent memory function is computed. The general memory function is expanded under the long wavelength limit and frequency dependent imaginary part of the memory function is calculated. The numerical computation of the real part of the memory function is performed using the Krammers-Kronig transformation. I found that real part conductivity shows the Drude peak at lower frequency and at higher frequency conductivity deviates from the Drude law and scales as $ \frac{1}{\om}$.
\section*{Appendix}
The terms for expression (\ref{fd12}) are
\begin{gather}
	h_{1}(\ep,\hbar\om) =\frac{\sqrt{\ep-\hbar\om}}{(1+e^{\beta(\ep-\hbar\om-\mus)})}-\frac{\sqrt{\ep+\hbar\om}}{(1+e^{\beta(\ep+\hbar\om-\mus)})}\\ 
	h_{2}(\ep,\hbar\om) =\frac{1}{\sqrt{\ep-\hbar\om}(1+e^{\beta(\ep-\hbar\om-\mus)})}-\frac{1}{\sqrt{\ep+\hbar\om}(1+e^{\beta(\ep+\hbar\om-\mus)})}\\~~~~~~~~~~~~~~-2\beta\frac{\sqrt{\ep-\hbar\om}e^{\beta(\ep-\hbar\om-\mus)}}{(1+e^{\beta(\ep-\hbar\om-\mus)})}+2\beta\frac{\sqrt{\ep+\hbar\om}e^{\beta(\ep+\hbar\om-\mus)}}{(1+e^{\beta(\ep+\hbar\om-\mus)})}\\ 
	h_{3}(\ep,\hbar\om)=\sqrt{\ep+\hbar\om}-\sqrt{\ep-\hbar\om},~~~~~~~h_{4}(\ep,\hbar\om)=\frac{1}{\sqrt{\ep+\hbar\om}}-\frac{1}{\sqrt{\ep-\hbar\om}} ,
\end{gather}
and $\Xi$ function are 
\begin{align}
	\Xi (\beta\mud)&=\int_{-\beta\mud}^{\infty}\frac{dx \sqrt{x+\beta\mud} e^{x}}{(e^{x}+1)^2}+\frac{2}{3}\int_{-\beta\mud}^{\infty}\frac{dx (x+\beta\mud)^{\frac{3}{2}} e^{x}}{(e^{x}+1)^2} -~\frac{4}{3}\int_{-\beta\mud}^{\infty}\frac{dx (x+\beta\mud)^{\frac{3}{2}} e^{2x}}{(e^{x}+1)^3} \\
	\Xi^{1}(\beta\mud)&=\int_{-\beta\mud}^{\infty}\frac{dx \sqrt{x+\beta\mud} e^{x}}{(e^{x}+1)^2}\\
	\Xi^{2}(\beta\mud)&= \int_{-\beta\mud}^{\infty}\frac{dx \sqrt{x+\beta\mud} e^{x}}{(e^{x}+1)^3}+\frac{2}{3}\int_{-\beta\mud}^{\infty}\frac{dx (x+\beta\mud)^{\frac{3}{2}} e^{x}}{(e^{x}+1)^3} -\frac{4}{3}\int_{-\beta\mud}^{\infty}\frac{dx (x+\beta\mud)^{\frac{3}{2}} e^{2x}}{(e^{x}+1)^4} .
\end{align}

\section*{Acknowlegement}
I thank Dr. Navinder Singh for encouragement and important comments.

	\end{document}